\documentclass[preprint,10pt]{elsarticle}
\usepackage[utf8]{inputenc}
\usepackage[T1]{fontenc} 
\usepackage{amssymb}
\usepackage{amsthm}
\usepackage{amsmath}
\usepackage{graphicx}
\usepackage{siunitx}
\usepackage{enumitem}
\usepackage{caption}
\usepackage{subcaption}
\usepackage{eufrak}
\usepackage{booktabs, multirow} 
\usepackage{soul}
\usepackage[table]{xcolor} 
\usepackage{changepage,threeparttable} 
\usepackage[colorlinks=true, allcolors=blue]{hyperref}
\usepackage{cite}
\usepackage{float}

\usepackage{lineno}

\newcommand{\replyTo}[1]{\textit{** Comment #1 ** }}
\renewcommand{\replyTo}[1]{}

\begin{document}

\begin{frontmatter}

\title{A novel methodology to estimate pileup effects and induced error in microdosimetric spectra}

\author[1,4]{E. Pierobon}
\ead{e.pierobon@gsi.de}
\author[2,4]{M. Missiaggia}
\ead{marta.missiaggia@miami.edu}
\author[3,4]{F. G. Cordoni}
\ead{francesco.cordoni@unitn.it}
\author[2,4]{C. La Tessa\corref{cor1}}
\ead{chiara.latessa@miami.edu}

\cortext[cor1]{Corresponding author}

\affiliation[1]{organization={GSI Helmholtzzentrum f\"ur Schwerionenforschung},
addressline={Planckstraße 1},
postcode={64291},
city={Darmstadt},
country={Germany}}

\affiliation[2]{organization={Department of Radiation Oncology, University of Miami Miller School of Medicine},
postcode={33136},
city={Miami, FL},
country={United States of America}}

\affiliation[3]{organization={Department of Civil, Environmental and Mechanical Engineering},
addressline={via Mesiano 77},
postcode={38123},
city={Trento},
country={Italy}}

\affiliation[4]{organization={Trento Institute for Fundamental Physics and Application (TIFPA)},
addressline={via Sommarive 15},
postcode={38123},
city={Trento},
country={Italy}}

\begin{abstract}
Microdosimetry provides a more granular characterization of the radiation field compared to conventional LET-based methodology, and for this reason has become increasingly attractive for quality assurance in particle therapy. However, the typical particle rates used in the treatment delivery lead to pileup, which distorts the experimental spectra, and thus compromises the accuracy of microdosimetric measurements, limiting their use in clinical settings.
In this work, we investigated the pileup effects in a commercial spherical Tissue Equivalent Proportional Counter (TEPC), and developed an algorithm to evaluate their impact on measured microdosimetric spectra. We exposed the TEPC to 11 and 70 MeV proton beams, and collected the microdosimetric spectra at rates in the $10^3$-$10^6 \, \unit{pps}$ range. Using a combination of Geant4 Monte Carlo simulations and experimental data, we developed an algorithm to evaluate the pileup probability affecting the experimental measurements.
\replyTo{1}The results indicate that, at a particle rate of $429 \pm 21$ pps, the pileup probability is negligible, with a value of $0.01 \pm 0.01 \,\%$.
Additionally, the comparison between the two proton energies suggests that the methodology proposed here to estimate pileup can be used to predict its effects in clinical proton beams with similar energies. Furthermore, the same approach can be applied to any type of microdosimetric measurement, making it a reliable tool for addressing pileup issues and facilitating the application of microdosimetry in clinical settings.
\end{abstract}

\begin{highlights}
\item Development of a stochastic algorithm to simulate pileup effect on simulated data.
\item Development of a methodology to estimate pileup contribution on experimental data.
\item To achieve an error below 5\% on the $y_F$, the pileup level has to be below $18 \pm 2 \, \%$, corresponding to a rate of $1.012 \pm 0.005\,  \times 10^4$ pps.
\item \replyTo{1}At a rate of $429 \pm 21$ pps, the pileup probability is negligible ($0.01 \pm 0.01\%$), resulting in an error on $y_F$ of less than $1\%$.
\end{highlights}


\end{frontmatter}

\clearpage
\section{Introduction}
Microdosimetry is an effective approach to describe the quality of the radiation field \citep{10.1093/rpd/ncq390, goodhead1982assessment}. 
Unlike linear energy transfer (LET), for which no standardized measurement methodology currently exists, microdosimetry provides a clear and operationally well-defined physical quantity: the lineal energy $y$ \citep{rossi1996microdosimetry, ICRU36, missiaggia2024radiation}. 
This unique strength is one of the reasons microdosimetry is increasingly recognized as a promising tool for clinical quality assurance (QA) \citep{MAGRIN2023109586, ROSENFELD2016156, bianchi2021repeatability, colautti2020characterizing}. Furthermore, the microdosimetric kinetic model (MKM), which is widely adopted in Asia and Europe for carbon ion therapy, uses microdosimetric quantities as input to describe radiation quality \citep{hawkins1996microdosimetric, sato2009biological, inaniwa2010treatment}.

The Tissue Equivalent Proportional Counter (TEPC) is considered the standard reference detector in microdosimetry \citep{ICRU36, MAGRIN2023109586, missiaggia2020microdosimetric, missiaggia2023investigation, colautti2020characterizing, missiaggia2024radiation}. TEPCs operate in a proportional regime, measuring energy loss by radiation at the micrometer level. By adjusting the gas pressure in the active region, it is possible to create a volume equivalent to only a few micrometers in size for a unit density material \citep{rossi1996microdosimetry}. 
Depending on the active volume size, TEPCs can sustain different particle rates, ranging from low intensities in the order \replyTo{3}of $10^3$ particles per second (pps) for spherical geometries of $12.6 \, \si{cm}$ diameter \citep{https://doi.org/10.1118/1.4916667}, to higher intensities of $ 10^{5} $ particles per second for cylindrical active regions with $ 0.9 \, \si{\milli m}$ of radius \citep{BIANCHI2024107271}. When the particle rate exceeds the detector capability, it leads to an effect known as pileup, which can distort the microdosimetric spectrum.

Obtaining high quality microdosimetric data without pileup is challenging, especially in clinical facilities where the particle rate is limited by design and operational constraints.
The primary limitation stems from the beam monitoring system, particularly the ionization chambers (ICs), which are typically used to measure the \replyTo{4}number of delivered particles, and therefore, the delivered dose.
The raw output signal of an IC is an electric current, which is typically not directly used. In most cases, electronic processing converts IC raw signals into a more suitable format, such as pulses, while maintaining the proportionality between beam intensity and the number of pulses. Counting these IC pulses (hereafter referred to as IC counts) is required for spot-scanning \citep{Pedroni_2005,HABERER1993296}, making IC an irreplaceable clinical component.
If anomalous measures are detected on the IC during the beam delivery (e.g., if the particle rate intensity is too low), the interlock system engages, halting the beam delivery.
Overcoming this constraint requires either detailed knowledge of the facility and sufficient flexibility to modify its operation, or access to dedicated experimental beamlines specifically designed for low-rate studies.
Because the IC readout cannot normally be bypassed in clinical facilities, lowering the particle rate to values that are acceptable for TEPC measurements is often not possible. 
Finally, current research is shifting toward FLASH applications with ultra-high dose rates, reducing the focus on the development of techniques to lower the particle rates \citep{romano2022ultra, nesteruk2021flash, jolly2020technical}.
Consequently, medical facilities are unable to lower particle rates to levels that would effectively mitigate the issue of pileup \citep{newhauser2009international}.

Pileup can originate from both the detector charge collection and the electronic acquisition chain \citep{knopf2025exploring}. If the particle rate is sufficiently high, exceeding the charge collection rate in the detector, an event can deposit charge in the active volume while the \replyTo{5}electric field is not quiescent. 
This accumulation eventually leads to an irregular proportionality between energy deposited and collected charge \citep{langen2002pileup}.
Pileup can also occur in the electronic acquisition chain if the time between \replyTo{6}consecutive events is too short and no effective pileup rejection is implemented \citep{langen2002pileup}.  The effect of electronic pileup depends on the sequential components of the chain, but typically causes the overlap of two or more signals, producing a distorted spectrum. \\
To mitigate pileup effects, different techniques have been developed over the years. A standard technique consists in modeling the system response as paralyzable or non-paralyzable \citep{knoll2010radiation}, which enables estimation of the events lost due to pileup through the system \textit{dead time} and prediction of the true detector count rate in the absence of pileup effects.
However, these corrections are applied exclusively to the total counts measured by the detector meaning that the effects on the microdosimetric spectra are not accounted for.
Pileup causes a broadening and a shift toward higher values of the microdosimetric lineal energy $f(y)$-distribution and, consequently, of all the derived distributions ($yf(y),\, d(y)$,\, $yd(y)$). This, in turn, results in an overestimate of the frequency-mean lineal energy $y_F$ and the dose-mean lineal energy $y_D$, both commonly used to characterize radiation quality in microdosimetry-based radiobiological modeling \citep{hawkins1994statistical, bellinzona2021linking}.  Inaccuracies in their estimation can therefore lead to poor predictive performance for biological outcomes.

Pileup in TEPCs has only been investigated with neutron beams and in a limited range of particle rates (3 kpps to 55 kpps) \citep{langen2002pileup}. This creates a significant gap in the field, as microdosimetry literature lacks pileup characterization at higher particle rates and with ion beams. Furthermore, no studies have assessed how pileup influences the microdosimetric spectra or the resulting errors in derived quantities such as $y_F$ and $y_D$. Notably, a recent study reported pileup effects in solid-state microdosimeters \citep{knopf2025exploring}, demonstrating that despite their substantially smaller cross-sectional area compared to TEPCs, this detector class still needs to actively counter the effects of pileup.

This work is organized into two parts. 
First, the response of a commercially available TEPC detector to particle rates ranging from approximately $10^3$ to $10^6$ \replyTo{7}pps is examined.
This, in turn, requires precise monitoring of the beam rate and estimation of the particle rate, and specific techniques developed for this purpose are presented. 
Furthermore, the system response to particles is characterized by its dead time, allowing the pileup effect to be assessed using a \replyTo{8}standard method based purely on counting statistics. We performed measurements at two proton energies: $70 \, \si{MeV}$  and $11 \, \si{MeV}$, which correspond to the entrance region \replyTo{9}(low LET) and peak (high LET) of the typical Bragg curve of protons, respectively. 
The second part of the work employs Monte Carlo simulations to quantify the contribution of pileup to the microdosimetric spectra as a function of the pileup probability, $p$.
Monte Carlo is particularly advantageous in this context because simulations are, by definition, free of pileup artifacts, and thus they provide an essential reference for identifying and characterizing pileup distortions observed in the experimental data. 
A novel algorithm that applies pileup effects to simulated spectra at a specified probability $p$ is presented. 
By combining experimental measurements with simulation-based references, we provide an accurate quantification of the impact of pileup on microdosimetric quantities.

Finally, the present work aims to introduce new effective tools for reproducing and quantifying the effects of pileup and its probability on experimental microdosimetric data. We demonstrate that, despite the presence of pileup, experimental microdosimetry can still provide valuable information for characterizing radiation quality in particle therapy.

\section{Material and methods}
\subsection{Experimental setup}
All measurements were performed at the physics beamline of the Proton Therapy Center in Trento (Italy) \citep{tommasino2017proton}.
The microdosimetric spectra were acquired using a commercially available spherical TEPC model LET-1/2 from Far West Technology. 
The latter was filled with propane gas, equivalent to a tissue sphere of $2\, \si{\micro m}$ in diameter, and was biased at $700 \, \si{V}$.
The beam was monitored with an Ionization Chamber (IC), manufactured by De.Tec.Tor, and of thickness equivalent to $\qty{0.6}{\mm}$ of water \citep{mirandola2018characterization, ROVITUSO2025104883}.
To calibrate the IC readout to an actual particle rate, a $1\, \unit{mm}$ thick BC400 plastic scintillator (SC) was deployed. 
As the SC can affect the radiation field, it was only used for calibrating the IC, and removed prior to acquiring the microdosimetric spectra.
Microdosimetric measurements and subsequent IC calibrations were conducted at two energies: $ 11 \, \si{MeV}$ and $70 \, \si{MeV}$. While the highest energy could be delivered directly from the facility, the $ 11 \, \si{MeV}$ energy was achieved by passively degrading the $70 \, \si{MeV}$ beam with $35\, \si{mm}$ of RW3 (solid water).  \\

\begin{figure*}[htb]
    \centering
     \begin{subfigure}[b]{0.49\textwidth}
         \centering
         \includegraphics[width=\textwidth]{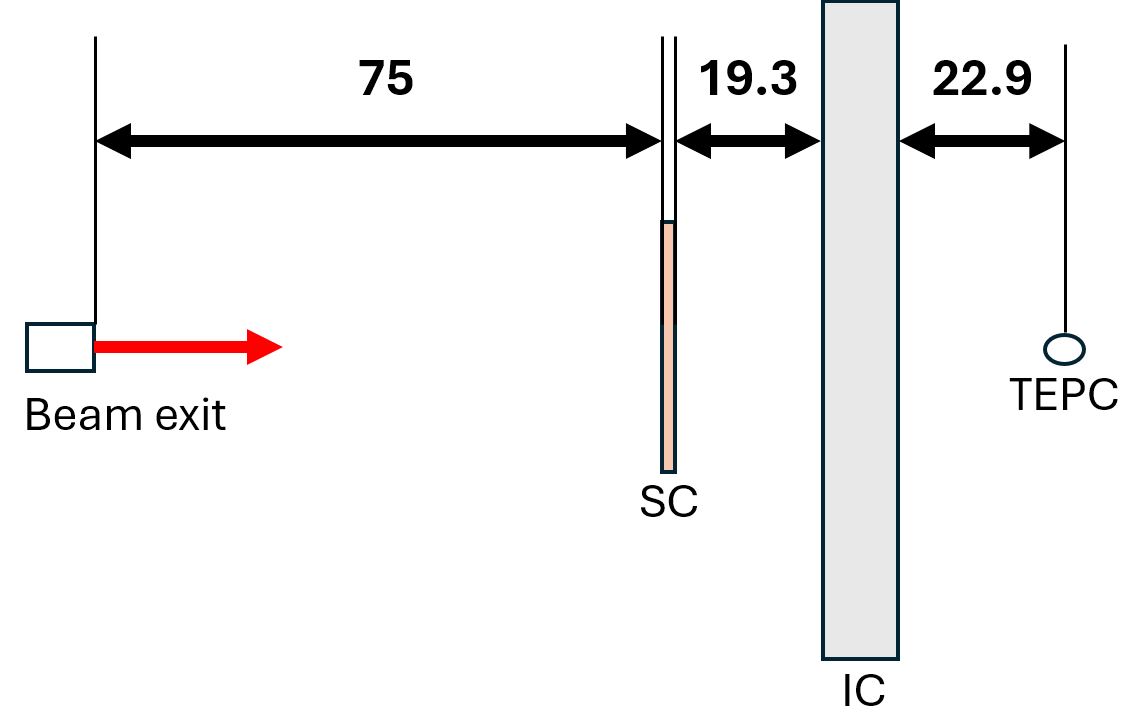}
         \caption{}
         \label{fig:exp_setup1}
     \end{subfigure}
     \hfill
     \begin{subfigure}[b]{0.49\textwidth}
         \centering
         \includegraphics[width=\textwidth]{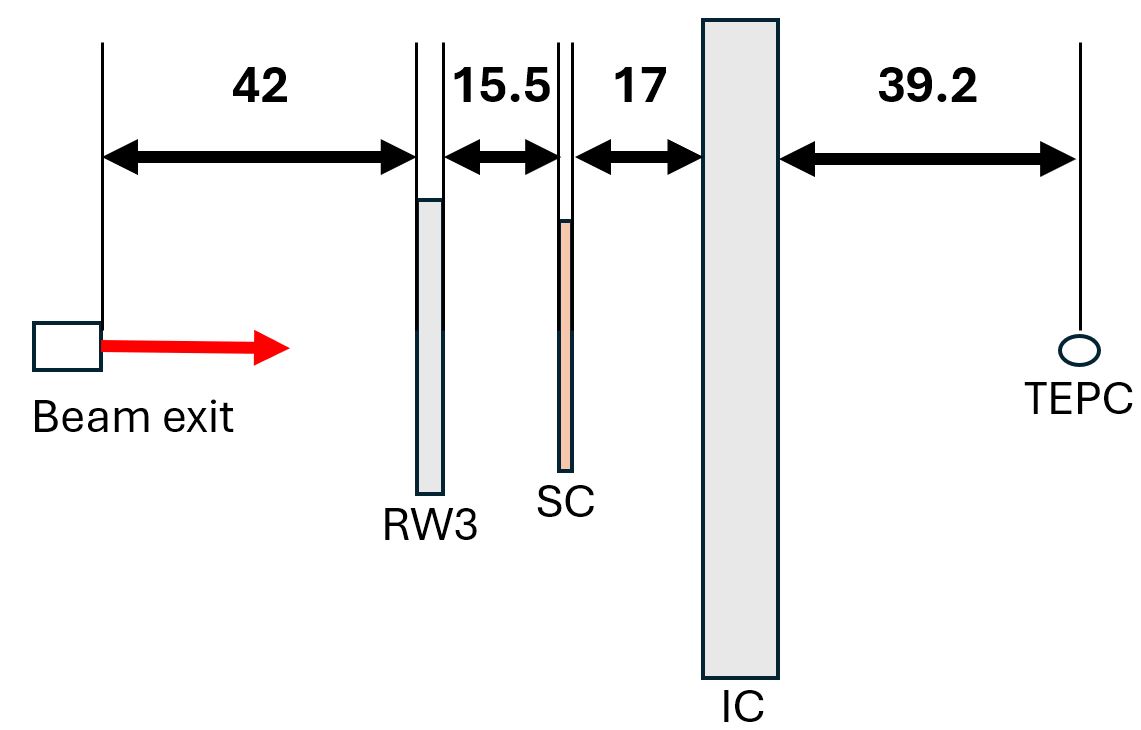}
         \caption{}
         \label{fig:exp_setup2}
     \end{subfigure}
     \hfill
    \caption{Experimental setup for the $70 \, \si{MeV}$ beam (left figure) and for the $11 \pm 5 \, \si{MeV}$ beam obtained by modulating the $70 \, \si{MeV}$ beam with $35 \, \si{mm}$ of RW3 (right figure). For both experiments, the scintillator (SC, $1\, \unit{mm}$ thick) was removed prior to the acquisition of the microdosimetric spectra. All distances are given in centimeters. The figure is to scale, with the exception of the TEPC active volume sphere for illustrative purposes. Ionization chamber is abbreviated as IC.}
    \label{fig:exp_setups}
\end{figure*}
The experimental setup for this study is shown in \autoref{fig:exp_setups}. 
The microdosimeter was always aligned in-beam, with its sensitive volume centered at the isocenter, defined $\qty{125}{\cm}$ away from the beam exit.
The distances between the elements were optimized to reduce secondary fragments generated by interactions of the beam with upstream elements. 
Suppression of these contributions is essential to avoid fragment contamination; experimental verification confirmed that the adopted distances are sufficient to prevent this effect in the microdosimetric spectra.
Due to the smaller cross-sectional area of the SC compared to the IC, the SC was always placed upstream of the IC. This was done to enhance the beam coverage of the SC, countering the beam's lateral divergence.
Finally, to ensure complete detector immersion within the beam profile, the beam dimensions were validated at the isocenter using Gafchromic film measurements for the 70 MeV beam energy, finding results consistent with \citep{tommasino2017proton}.
For the 11 MeV beam energy, it can be reasonably assumed that the lateral beam divergence induced by the RW3 material further broadened the beam profile.
\subsection{Electronic acquisition chain}
The readout chain of the SC consists of a CAEN discriminator module N841 and the number of detected pulses was counted with a timer and counter module ORTEC-871. 
The IC readout system provides IC-counts, which can be directly related to the beam intensity. The full readout system, composed of both proprietary hardware and software, is not publicly available as the system is commercial. The software allows the final user to apply calibrations to directly convert the frequency of the IC-counts to a particle rate (IC particle rate hereafter) depending on the configuration/energy. 
Ultimately, the software generates files containing IC counts and relative timestamps with a resolution of $200 \, \unit{m\s}$, enabling real-time monitoring and assessment of the beam temporal uniformity.

The electronic acquisition chain of the TEPC included a CAEN A422A charge-sensitive preamplifier as the first stage of the readout. To guarantee that each signal was processed independently of the amplitude throughout the dynamic range while maintaining an adequate resolution, after the preamplification stage the signals were fed into three shaping amplifiers: two CAEN N968 models with gains of $1000$ and $100$ (referred to as high and medium gain, respectively) and an Intertechnique 7243E model with a gain of $10$ (referred to as low gain). The shaping time constant of all amplifiers was set at $2 \, \si{\micro s}$, as the optimal compromise between electronic noise, signal integrity and signal duration. 
Additionally, the $2\,\si{\micro s}$ allows for a controlled exposure of the system to pileup for investigating its effects.
The amplifier gains were selected to have an overlap region between the high and medium spectra, and between the medium and low spectra. This method allows to merge the three spectra into one in post-processing, covering the entire dynamic range with sufficient resolution. The signals generated by the three amplifiers were sent to a multichannel analyzer (MCA). The medium and high gain pulses were fed into an MCA model 927 by ORTEC, while the low gain signal was sent to an MCA model 926 by ORTEC. The MCAs were read out with the MAESTRO software, which generates files containing the counts recorded in each channel. The MCAs can be equipped with a pileup rejection mechanism, which was intentionally disabled to study the effects of pileup on the system without introducing external bias.

The MAESTRO software measures both the raw acquisition time (real time $t_r$) and the time adjusted for the acquisition and processing time (live time $t_l$). 
These two quantities account for the fact that MCA typically requires a certain amount of time to acquire each pulse, referred to as \replyTo{10}non-extendable dead time ($t_r - t_l$) \citep{rovituso2017fragmentation, knoll2010radiation}. During dead time, the acquisition cannot process any other signal, and thus all events occurring within this interval are lost. 
To compensate for event losses, the dead time for each amplification was measured, and the corresponding MCA counts were corrected by multiplying them by the ratio of real time to live time. TEPC counts or rates corrected by live time are referred to TEPC$_{\text{LTC}}$. On the other hand, TEPC$_{\text{RAW}}$ refers to the counts or rates where no correction is applied.

\subsection{Monte Carlo simulations}
\label{sec:MM_MC_SIM}
\begin{figure}
    \centering
    \includegraphics[width=\textwidth]{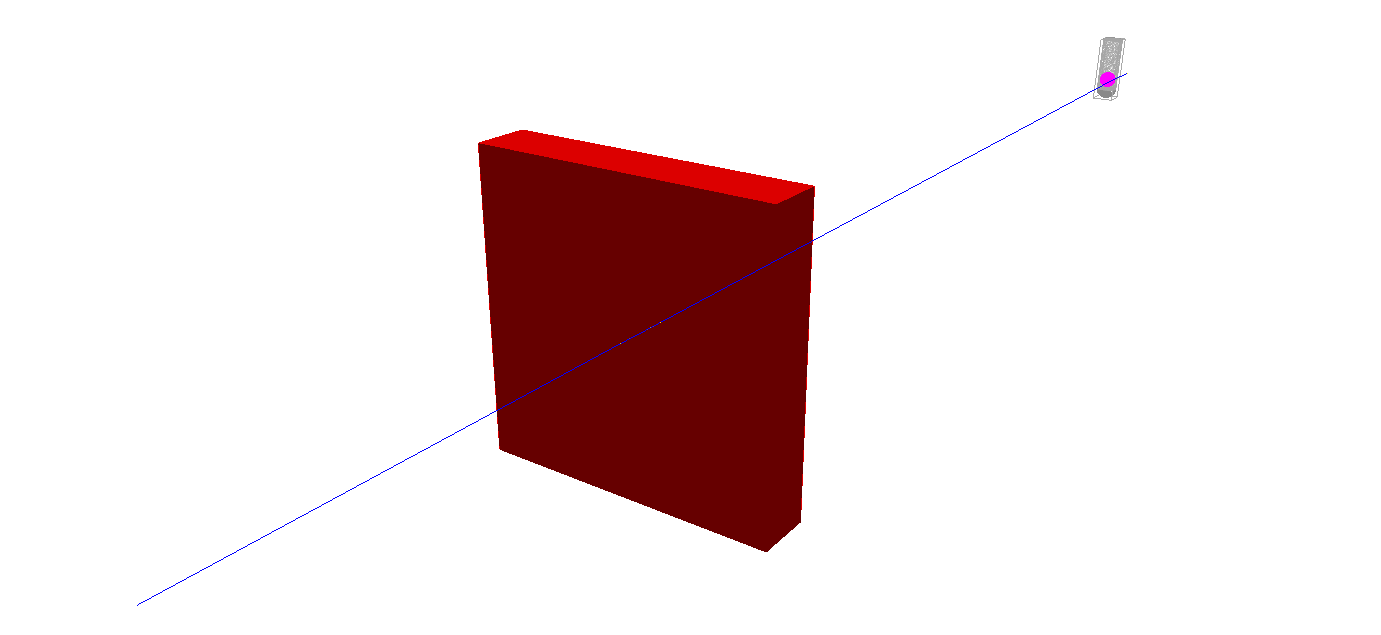}
    \caption{Picture of the $\qty{11}{\mega \eV}$ setup implementation in Geant4 Monte Carlo toolkit. The beam originates at the edge of the world, on the left side of the picture. The $\qty{37}{\milli \m}$ of RW3 water equivalent material is shown in red. The TEPC detector is visible on the far right of the picture, 125 cm away from the beam origin, with its active volume highlighted in purple. The energy deposited by all particles was scored in the TEPC active volume. A proton track is visible in blue, traversing left to right and reaching the sensitive volume of the TEPC.}
    \label{fig:MC_sim}
\end{figure}

All simulations were performed using Geant4 Monte Carlo toolkit \citep{1610988, AGOSTINELLI2003250, ALLISON2016186} version 10.06.p01, implementing the experimental geometry illustrated in \autoref{fig:exp_setups}. The single scattering mode was used to describe electromagnetic interactions, while hadronic interactions were described by the high-precision G4\_QGSP\_BIC\_HP physics list. \\
A picture of the Geant4 simulation setup for the $\qty{11}{\mega \eV}$ energy is shown in \autoref{fig:MC_sim}. 
A better agreement between the experimental and simulated microdosimetric spectra, both in terms of shape and in peak position, was achieved for the $\qty{11}{\mega \eV}$ energy by introducing an additional $\qty{2}{\milli \m}$ of RW3 water equivalent material.
For this reason, $\qty{37}{\milli \m}$ of RW3 were considered in all simulations featuring the $\qty{11}{\mega \eV}$ energy instead of the $\qty{35}{\milli \m}$ used experimentally.
The IC was not modeled in the Monte Carlo setup due to its negligible water equivalent thickness. \\
The TEPC detector was implemented as described in \citep{missiaggia2021novel,PIEROBON202683}.  
The beam parameters (energy, energy spread and dimensions) were selected and modeled according to the characterization provided in \citep{tommasino2017proton}.

\section{Results}
In this study, the link between particle rate and pileup was examined for a commercial spherical TEPC exposed to proton beams at different therapeutic energies. This process entails two fundamental components: the experimental microdosimetric distribution, measured at a well-defined particle rate, and the simulated microdosimetric distributions, generated at a specified pileup probability. 
The experimental and simulated microdosimetric spectra are then compared using an objective metric, to establish a relationship between particle rate and pileup probability.
An additional commonly adopted method to estimate pileup effects based on dead-time is applied to the dataset with the aim of validating our proposed approach.

\subsection{Beam monitoring and particle rate estimation}
\label{sec:results:exp_pileup}
In principle, the event rate at the TEPC can be estimated with the MAESTRO software. 
However, the counts from MAESTRO are affected by pileup distortions that become more severe as the particle rate increases.
To precisely determine the particle rate incident on the TEPC across a wide range of beam intensities, the IC measurements were used.
By comparing the particle rate measured by the IC with the event rate determined by MAESTRO using the corrected live time counts in the region where pileup distortion is not severe, a linear relationship can be established.
This calibration enables a robust extrapolation of the particle rate at the TEPC (hereafter TEPC$_{\text{IC}-\text{SC}}$) across the full dynamic range accessible to the IC.

\begin{figure*}[h]
    \centering
     \begin{subfigure}[b]{0.49\textwidth}
         \centering
         \includegraphics[width=\textwidth]{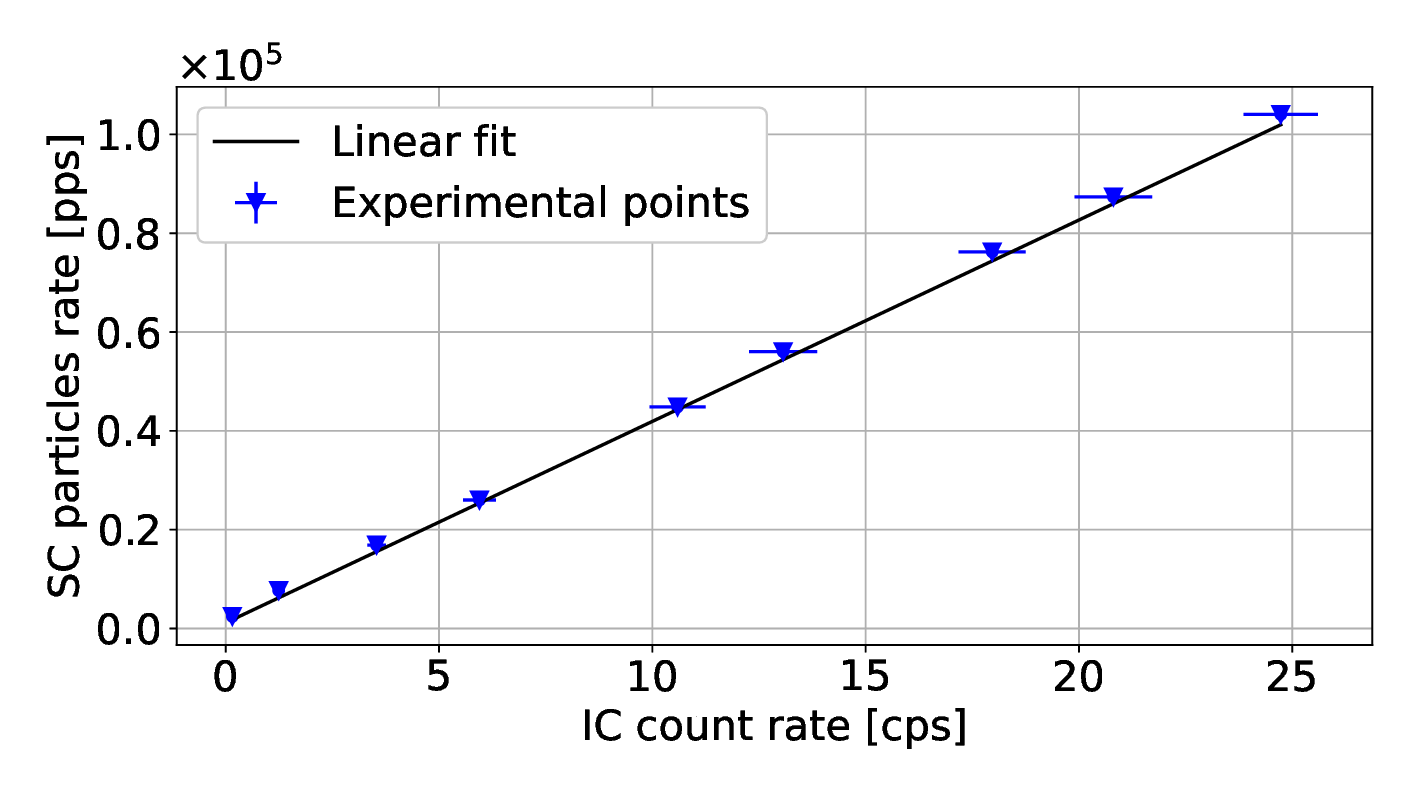}
         \caption{$70\, \si{MeV}$ proton energy.}
         \label{fig:calib_70_IC_vs_SC}
     \end{subfigure}
     \hfill
     \begin{subfigure}[b]{0.49\textwidth}
         \centering
         \includegraphics[width=\textwidth]{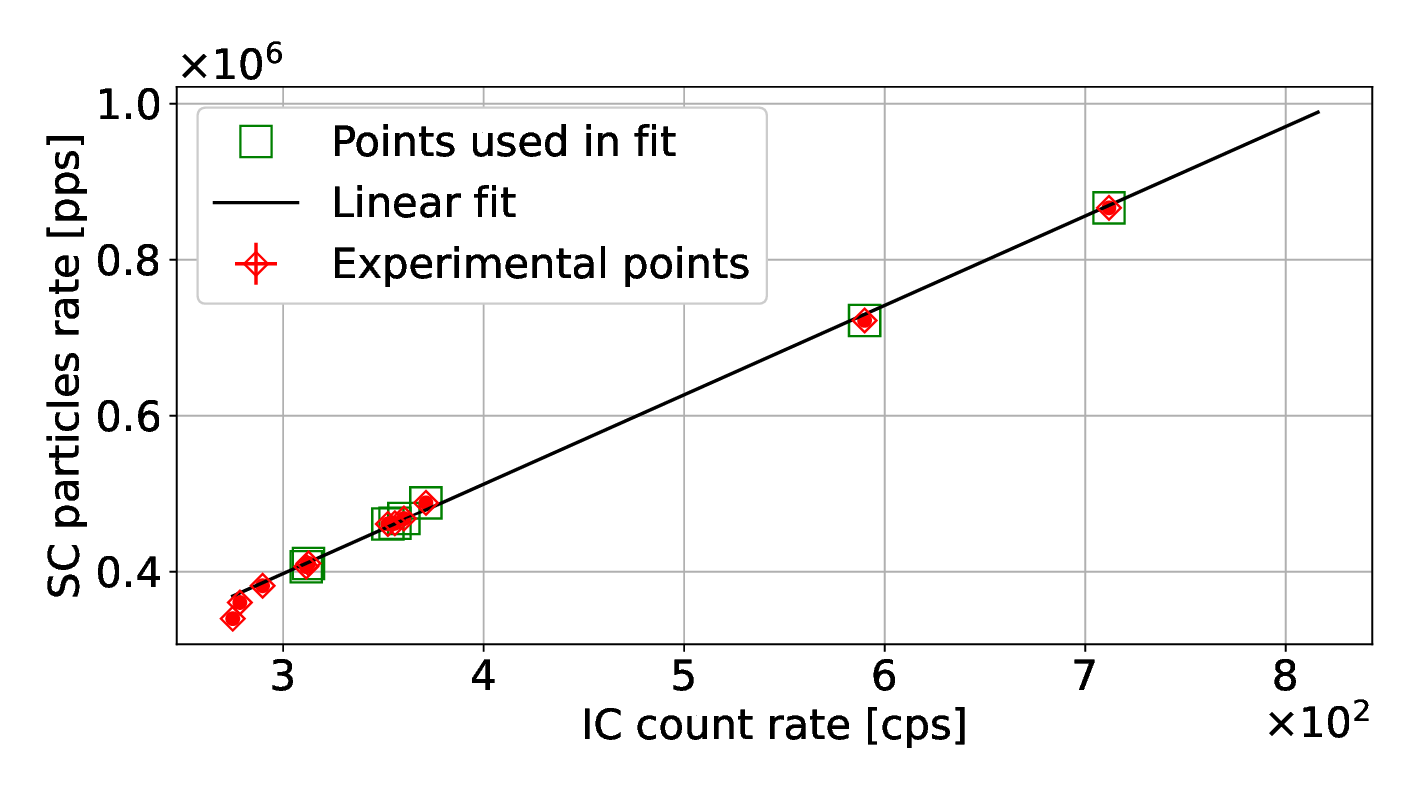}
         \caption{$11\, \si{MeV}$ proton energy.}
         \label{fig:calib_mod_IC_vs_SC}
     \end{subfigure}
     \hfill
    \caption{Particle rate measured with the SC versus IC count rate acquired for the two proton energies of 70 and 11 $\si{MeV}$. The experimental points (red and blue markers) were fit with a linear function, represented by black solid lines. While for the $70\, \si{MeV}$ dataset all the available data was used in the fit, for $11\, \si{MeV}$ only the data marked with a green square was used. The linear fit parameters are reported in \autoref{tab:chi2_IC_SC}. Uncertainties were estimated by propagating the statistical uncertainties.}
    \label{fig:IC_vs_SC_calib}
\end{figure*}

As the IC does not directly measure the particle rate but rather a count rate, the initial step in the proposed approach is to convert the IC count rate into a particle rate.
The challenge was addressed by deploying a plastic scintillator to perform subsequent linear calibration.
\autoref{fig:IC_vs_SC_calib} shows the particle rate measured at the SC \replyTo{11}as a function of the IC count rate, both acquired under the same beam conditions.

While the experimental dataset collected at $70 \, \si{MeV}$ exhibits a consistent linear behavior across the entire range (\autoref{fig:calib_70_IC_vs_SC}), at $11 \, \si{MeV}$, the IC begins to feature a non-linear response when the SC rate drops below $ 4 \times 10^5 \, \unit{pps}$ (\autoref{fig:calib_mod_IC_vs_SC}). 
We hypothesized that this deviation stems from two factors. First, slower protons release more energy in the IC, generating a larger signal. Second, the absorber that reduces proton energy to 11 MeV causes the beam spot to widen due to electromagnetic and nuclear scattering. 
Consequently, more protons are deflected at angles large enough to miss the IC. Therefore, the particle rate is reduced in the 11 MeV setup compared to the 70 MeV setup if the same intensity is required at the cyclotron. However, the particles reaching the IC release more energy, generating a higher signal. 
The net effect is that, at a low particle rate, more specifically, when the SC rate is below $4 \times 10^5 \,\unit{pps}$, the IC response is nonlinear in the 11 MeV setup. 
Therefore, the experimental points collected at $11 \, \si{MeV}$ with an SC rate below $4 \times 10^5 \, \unit{pps}$ were excluded from the IC–SC fit, as they exhibited a clear deviation from linear behavior.
Using the function $y = mx + q$, both the $11 \, \si{MeV}$ and $70 \, \si{MeV}$ datasets were fitted to obtain the black linear function plotted in \autoref{fig:IC_vs_SC_calib}, and whose parameters are reported in \autoref{tab:chi2_IC_SC}. 
The $m$ coefficient increases with proton energy, reflecting the fact that low-energy protons deposit more energy, leading to a higher IC signal and, consequently a lower calibration slope. 

\begin{table}[htb]
\centering
\scriptsize
\begin{tabular}{lrrrrr}
Energy & $m \, \left[\unit{pps}\right]$ & $\delta m \, \left[\unit{pps}\right]$ &$q \, \left[\unit{pps}\right]$ &$\delta q \, \left[\unit{pps}\right]$
\\\cmidrule{1-5}
$11\, \si{MeV}$  &$1146$ & $267$ & $53901$ & $9421$ \\\cmidrule{1-5}
$70\, \si{MeV}$  & $4277$ & $54$ & $2355$ & $582$ \\\cmidrule{1-5}
\end{tabular}
\caption{Calibration coefficients obtained from the linear regressions of \autoref{fig:IC_vs_SC_calib} using the $y = mx +q$ equation. Uncertainties, denoted by $\delta$, are derived from the errors associated with the linear regression.} 
\label{tab:chi2_IC_SC}
\end{table}

\begin{figure*}[htb]
    \centering
     \begin{subfigure}[b]{0.49\textwidth}
         \centering
         \includegraphics[width=\textwidth]{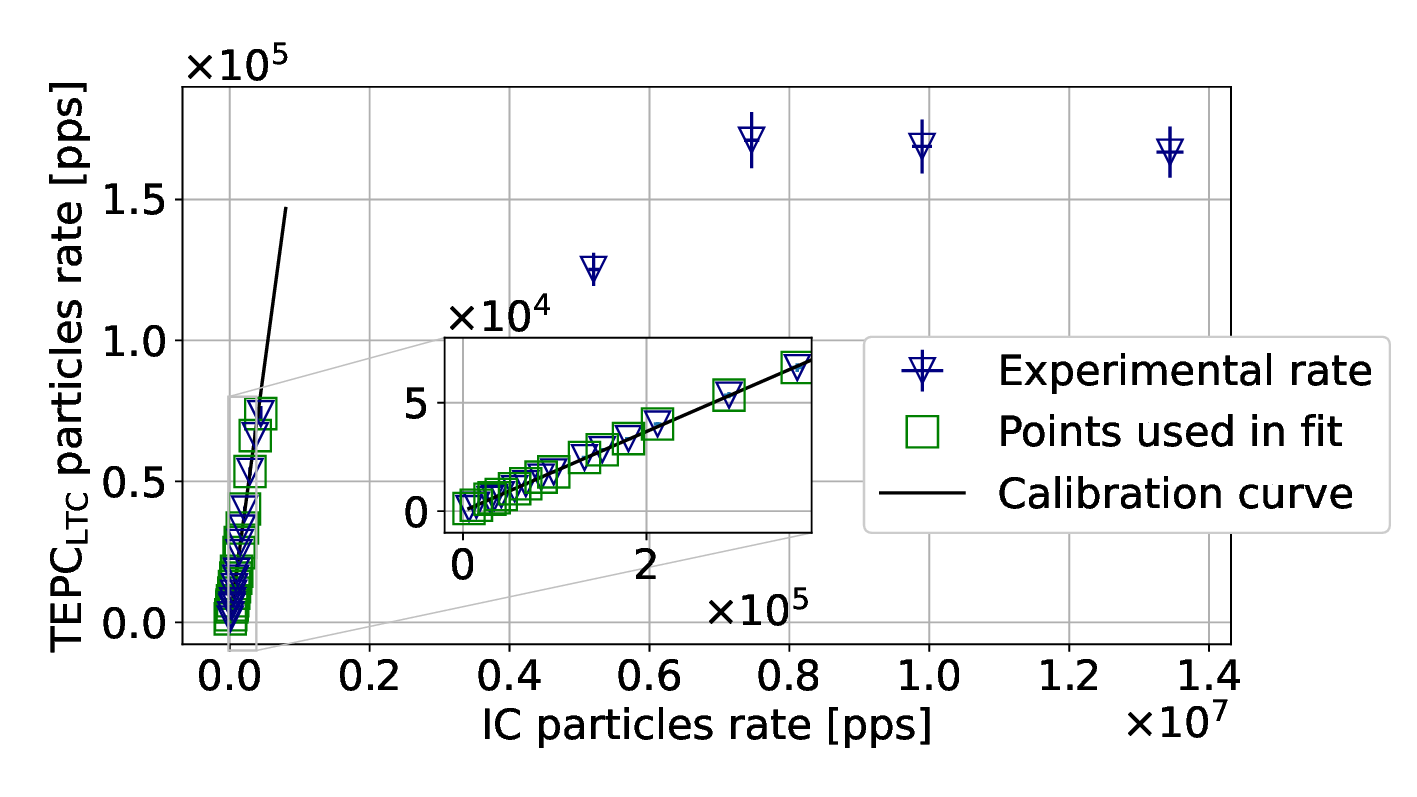}
         \caption{$70\, \si{MeV}$ proton energy.}
         \label{fig:calib_70_IC_vs_TEPC}
     \end{subfigure}
     \hfill
     \begin{subfigure}[b]{0.49\textwidth}
         \centering
         \includegraphics[width=\textwidth]{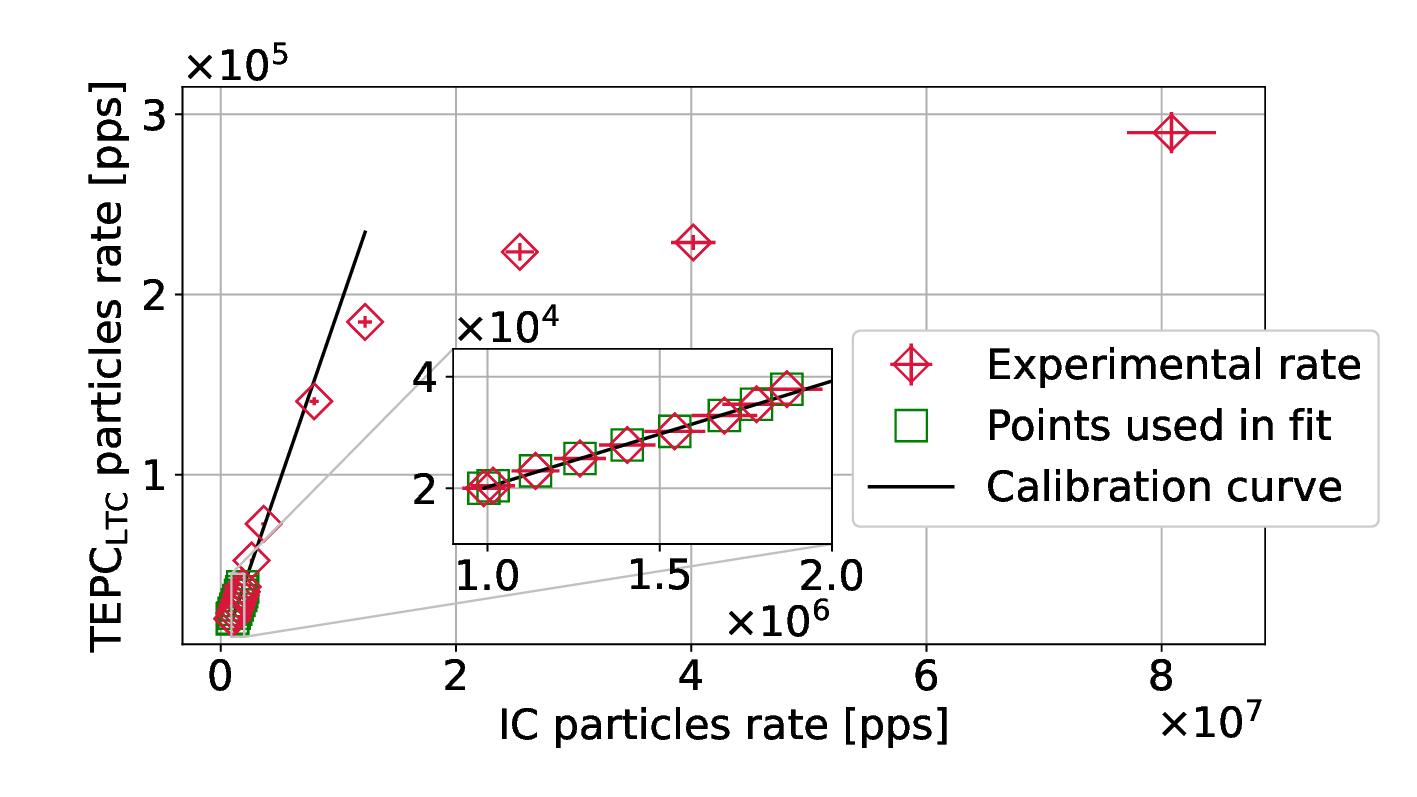}
         \caption{$11\, \si{MeV}$ proton energy.}
         \label{fig:calib_mod_IC_vs_TEPC}
     \end{subfigure}
     \hfill
    \caption{Live-time corrected particle rates at the TEPC (TEPC$_{\text{LTC}}$) versus calibrated ionization chamber (IC) measured for the two beam energies. A zoom of the low-rate region is shown for both graphs in the box. Points marked by the green square were used in the linear fit. The black line in the graph represents the best fit of the equation $y = mx + q$ to the selected dataset. The parameters of this equation are reported in \autoref{tab:calib_mod_IC_vs_TEPC}. Uncertainties were estimated by propagating the statistical uncertainties of the measured quantities.
    }
    \label{fig:IC_vs_TEPC_calib}
\end{figure*}

Both curves are not compatible with the $(0, \, 0)$ point (zero-rate point) within the estimated errors suggesting that a full line equation $y = mx + q$, which includes an offset parameter, is required to precisely calibrate the IC counts. \\
After converting the IC counts into a particle rate using the calibration in \autoref{tab:chi2_IC_SC}, these values can be directly compared with the MAESTRO readings. \autoref{fig:IC_vs_TEPC_calib} presents this comparison across different beam currents.
At both proton energies, the TEPC$_{\text{LTC}}$ rate is proportional to the IC below $\sim 5 \times 10^4 \, \unit{pps}$, while it saturates above. 
This saturation effect is likely to be induced by the acquisition chain when the particle rate exceeds its capability. 
To avoid any biasing and artifact, with the goal of obtaining the best possible particle rate estimation, only data in the linear region, marked by a green square in \autoref{fig:IC_vs_TEPC_calib}, was used in the fit. 
It is important to emphasize that, by restricting the data points considered within this interval, the applied live-time correction is correspondingly limited to those selected points. 
A linear function $y = mx+q$ was used, obtaining the calibration coefficients listed in \autoref{tab:calib_mod_IC_vs_TEPC}.
The black line in \autoref{fig:IC_vs_TEPC_calib} represents the aforementioned fit, and can therefore be interpreted as the predicted particle rate at the microdosimeter in the absence of pileup counting inefficiencies.

\begin{table}[H]\centering
\scriptsize
\begin{tabular}{lrrrrr}
Energy & $m$ &$\delta m$ &$q$ &$\delta q$ &  \\\cmidrule{1-5}
$11\, \si{MeV}$    & $0.0190$  & $0.0004 $ & $1199$ & $471$ \\
$70\, \si{MeV}$    & $0.1774$    & $0.0048 $  & $392$ & $3$ \\\cmidrule{1-5}
\cmidrule{1-5}
\end{tabular}
\caption{Calibration coefficients obtained from a linear regression of the TEPC$_{\text{LTC}}$ rate and the IC in \autoref{fig:IC_vs_TEPC_calib}. Uncertainties, denoted by $\delta$, are derived from the errors associated with the linear regression.} 
 \label{tab:calib_mod_IC_vs_TEPC}
\end{table}

The $m$ values associated with the two beam energies differ by approximately one order of magnitude. 
This discrepancy suggests that a significantly larger number of particles will traverse the IC but not the TEPC at $11 \, \si{MeV}$ energy, due to lateral scattering along their trajectory. 
As for the IC-SC calibration, the intercept of both curves is not zero, indicating that a full line equation $y = mx + q$ is necessary for an accurate description.
Finally, this calibration enables the estimation of the TEPC$_{\text{IC}-\text{SC}}$ rate, i.e. the particle rate at the detector without saturation effects.

\subsection{Pileup estimation based on Monte Carlo}
\label{sec:res:pileup_estimation_MC}
Based on the setup described in \autoref{sec:MM_MC_SIM}, we performed Monte Carlo (MC) simulations of the experimental beamline. 
Each event releasing energy in the detector was scored, obtaining a dataset of energy deposition events.
In the absence of pileup effects, the events are used independently to create microdosimetric spectra. 
\replyTo{14}Pileup effects were modeled by assuming that the number of additional interactions follows a Poisson distribution with mean $\lambda$. The corresponding pileup probability (i.e., the probability that an event is affected by pileup) is then given by:
\begin{equation}
P_{\mathrm{pileup}} = 1 - e^{-\lambda}. \label{eq:pileup_poisson}
\end{equation}
For a prescribed pileup probability, the parameter $\lambda$ was obtained by numerically inverting \autoref{eq:pileup_poisson}. The multiplicity $k$, defined as the number of additional pileup interactions associated with a given event, was sampled from a Poisson distribution with mean $\lambda$ using the \texttt{R} function \texttt{rpois}.

\begin{figure}[htb]
    \centering
    \includegraphics[width=1\textwidth]{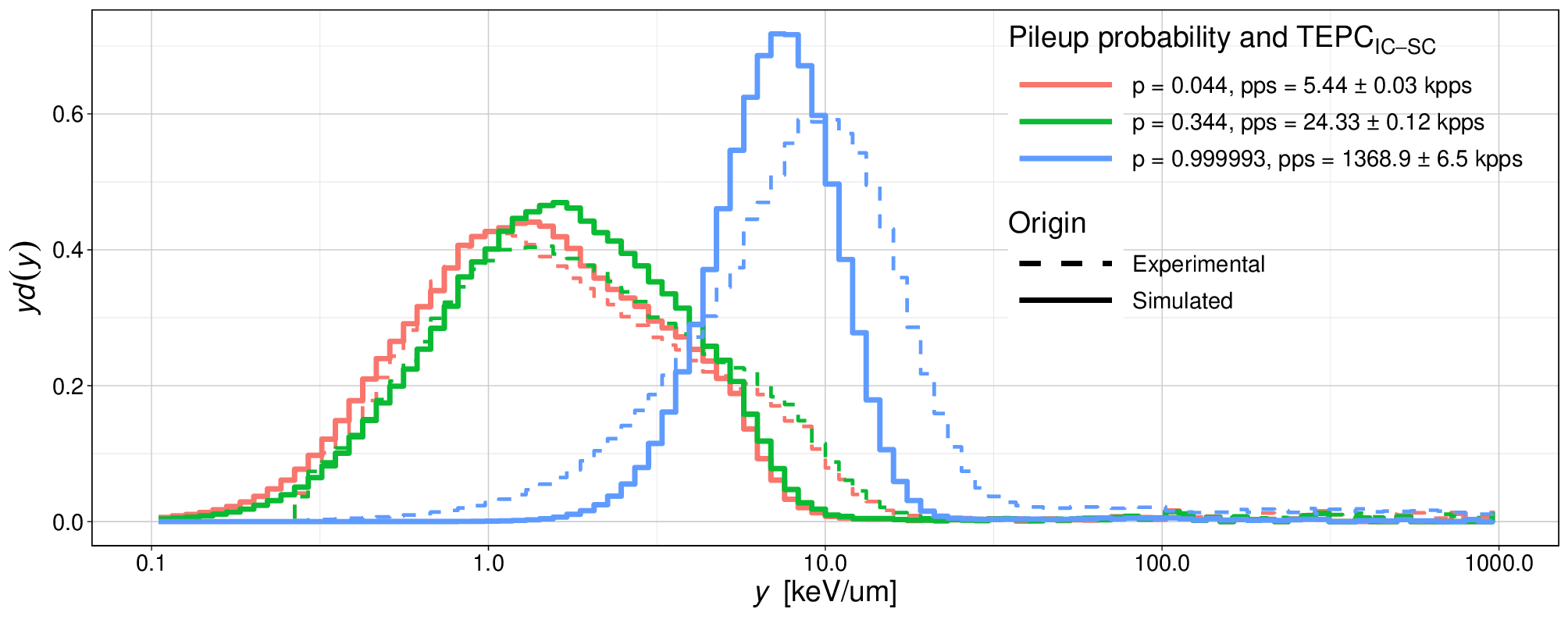}
    \caption{Simulated (thick dotted line) microdosimetric $yd(y)$ spectra of $70 \, \si{MeV}$ protons energy with different pileup probability $p = 0.044, \, 0.344$ and $0.999993$. Using the K-S test, it has been found that the experimental spectra that best match this pileup were acquired at a corrected particle rate (TEPC$_{\text{IC}-\text{SC}}$) of $5.44 \pm 0.03 $, $24.33 \pm 0.12$ and $1368.9 \pm 6.5$ kpps, respectively (continuous line).}
    \label{fig:Pile_up_sample_70MeV}
\end{figure}

More formally, a set of user-defined pileup probabilities $\left\{ p_{0}, \dots, p_{N} \right\}$ was defined, ranging from $0.001$ to $0.9999999$. To each probability $p_i \in \left\{ p_{0}, \dots, p_{N} \right\}$, the corresponding parameter $\lambda_i$ was calculated. At a fixed $\lambda_i$, for each energy deposition event $\epsilon_{0}, \dots, \epsilon_{M}$, a multiplicity $k_{0}, \dots, k_{M}$ was drawn. If  $ \left\{ k_0 \, \dots k_M \right\} \ni k_j = 0$, no action was performed.
If $ k_j \geq 1$, \replyTo{15}$k_j$ energy deposition events were randomly sampled from the reference dataset and summed to the original event. Formally, if $\epsilon_j$ denotes the original energy deposition associated with multiplicity $k_j$, the resulting energy deposition is given by:
\begin{equation}
    E_j = \epsilon_j + \sum_{l=1}^{k_j} w_l \, \epsilon_l,
\end{equation}
where the weights $w_l$ were independently sampled from a uniform distribution, $w_l \sim \mathcal{U}(0,\,1)$. The original energy deposition $\epsilon_j$ was retained with unit weight to preserve the baseline contribution. 
These sets $\left\{ E_0, \dots E_M \right\}_i$ were then used to calculate the microdosimetric spectrum at pileup probability $p_i$.
The process is repeated for 999 pileup probabilities from $0.001$ to $0.9999999$, obtaining a comprehensive set of microdosimetric spectra for both the $11 \, \si{MeV}$ and $70 \, \si{MeV}$ proton energies. 
An example of the algorithm effect is shown in \autoref{fig:Pile_up_sample_70MeV} for the $70 \, \si{MeV}$ beam energy. 
The dotted lines in various colors represent the  $yd(y)$ microdosimetric spectra obtained at various pileup probabilities ($p = 0.044, 0.344,$ and $0.999993$). As pileup probability increases, it is more likely that the energy deposition recorded for an event originates from the sum of multiple particles rather than from a single particle. 
The direct effect is that higher energy depositions become more probable, and, consequently, the microdosimetric spectrum shifts to larger $y$ values, as observed in \autoref{fig:Pile_up_sample_70MeV}.

The simulated microdosimetric measurements with varying pileup probabilities were then compared to the experimental measurements at different particle rates, aiming to establish a relationship between pileup probability and TEPC$_{\text{IC}-\text{SC}}$ particle rate.
To measure the agreement between the simulated and experimental spectra, we employed the Kolmogorov–Smirnov (K-S) test \citep{eadie1971statistical}. This nonparametric test evaluates the likelihood that two probability density distributions satisfy the null hypothesis of being statistically equivalent. 
A distinguishing feature of this approach is its capacity to provide an objective and reliable quantification of the similarity between two probability distributions. 
This offers an intuitive interpretation of the test as a measure of how closely the distributions align.

The K-S test was performed on all simulated–experimental microdosimetric $d(y)$ spectra pairs, considering a $\mathfrak{p}$-value above $95 \%$ to be statistically significant. To analyze the data, a matrix ($SM$) was defined according to \autoref{eq:sim_matrix_KB}, where the entry at position ($i,\,j$) is the value of K-S test ($D_{\text{K-S}}$) evaluated at the $i$-th experimental spectrum against the $j$-th simulated spectrum:
\begin{equation}
\begin{aligned}
    SM :=  \left\{ D_{\text{K-S}} (d(y)_i, \, d(y)_j) \right\}_{i j}, \, & i \in \{\text{experimental spectra}\}, \\
    & j \in \{\text{simulated spectra}\}
    \label{eq:sim_matrix_KB}
\end{aligned}
\end{equation}

The test was conducted within a subset of the most probable interval of the $d(y)$ spectrum, which accounts for at least 80\% of the spectrum counts, to enhance its robustness.
This region is mostly populated by primary protons, whereas the tails of the spectrum are more likely to contain rare secondary particles. By setting a $\mathfrak{p}$-value threshold above $95\%$, multiple simulated spectra with different pileup probabilities may be found compatible with a given experimental spectrum according to the K–S test. 
In this case, the lowest and highest probabilities among the compatible simulations were used to establish the uncertainty interval for the pileup range associated with the specific experimental spectrum.
\autoref{fig:Pile_up_sample_70MeV} shows the experimental spectra (continuous line) in best agreement with the simulated spectra (dotted lines) according to the K-S test for different pileup probabilities (different colors). 

\begin{figure}[htb]
    \centering
    \includegraphics[width=1\textwidth]{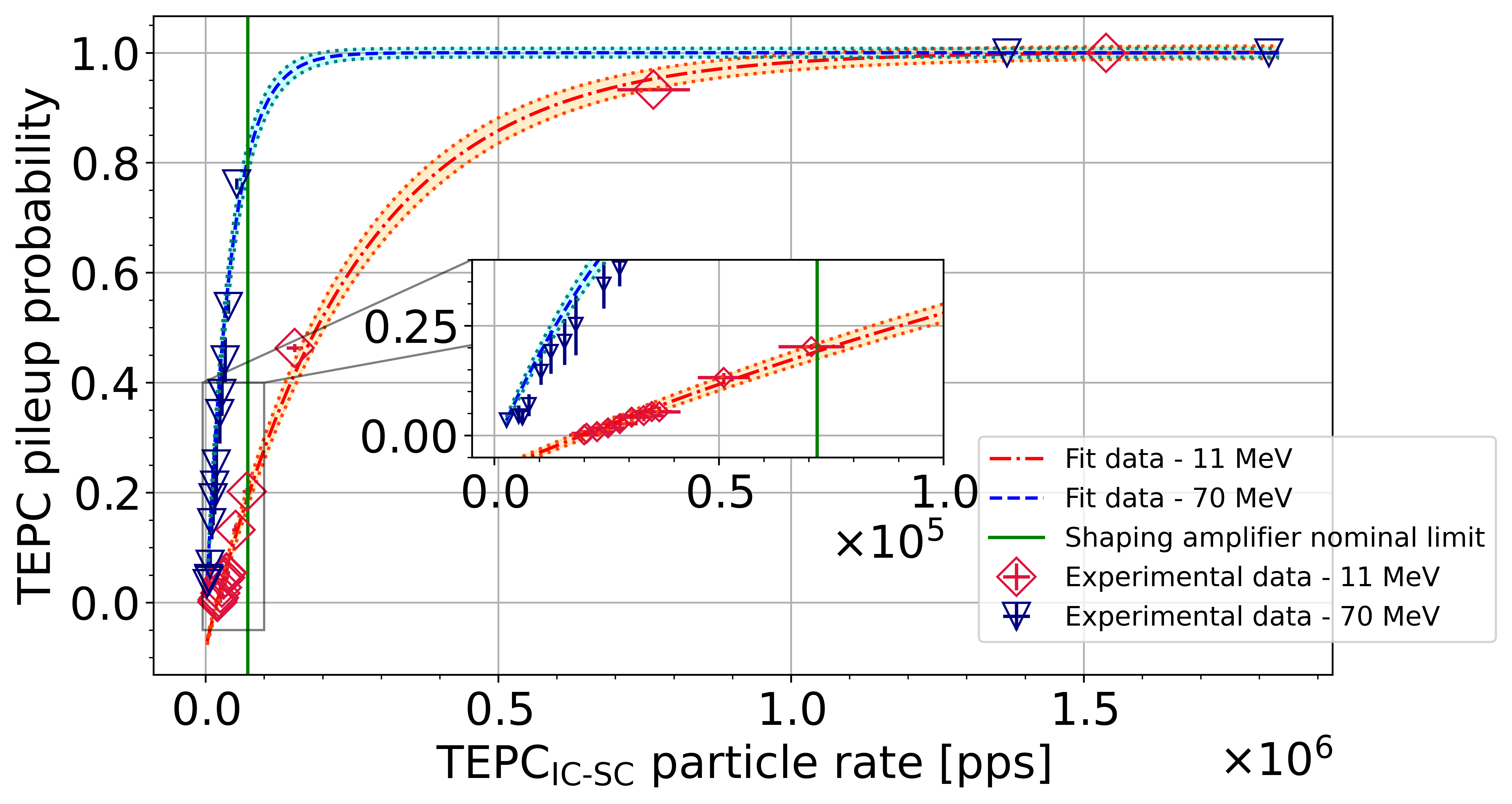}
    \caption{Pileup probability as a function of the particle rate at the TEPC$_{\text{IC}-\text{SC}}$ for $11 \, \si{MeV}$  (red diamonds) and $70 \, \si{MeV}$ (blue triangles) proton beams. 
    The reported uncertainties are derived from the error propagation of the statistical error.
    The dashed lines represent the fit obtained using \autoref{eq:pile_up_fit}, while the shaded region around each line denotes the corresponding confidence interval.
    The inverse of the typical signal duration according to the shaping amplifier manual of $(2.9 \cdot 2.4 \cdot t_{\text{shaping}})^{-1}$ is shown as a vertical green line.}
    \label{fig:Pile_up_vs_rate}
\end{figure}

By applying the K-S test to the entire dataset available, we established a relationship between the experimental and simulated spectra. 
This, in turn, directly enabled the correlation between the corrected particle rate at the TEPC (TEPC$_{\text{IC} - \text{SC}}$) and pileup probability.
\autoref{fig:Pile_up_vs_rate} shows the correspondence found between particle rate and pileup probability for the two energies of $11 \, \si{MeV}$ and $70 \, \si{MeV}$. Each point of the plot corresponds to a match found by the K-S test. Microdosimetric spectra exhibiting severe pileup distortion were automatically excluded based on the K–S test due to their substantial deformation from pileup effects.
The pileup-rate curve of \autoref{fig:Pile_up_vs_rate} can be modeled with the following equation: 
\begin{equation}
    p(x) = a + b \left( 1 - \exp \left[ - \frac{x}{\mu} \right] \right)\, ,
    \label{eq:pile_up_fit}
\end{equation}
which we used to fit the TEPC$_{\text{IC}-\text{SC}}$ data points of both energy datasets. The derived fit parameters are provided in \autoref{tab:fit_exp_pileup}.
\begin{table*}[!htb]
\centering
\scriptsize
\begin{tabular}{lrrrrrrrr}
Energy &$a$ &$\delta a$ &$b$ &$\delta b$ &$\mu \left[ \unit{pps} \right]$ &$\delta \mu \left[ \unit{pps} \right]$ & $\chi^2_r$ \\\cmidrule{1-8}
$11\, \si{MeV}$  & $-0.082$ &  $0.006$ & $1.084$ & $0.005$ & $247121$ & $10853$  &$2.2$ \\\cmidrule{1-8}
$70\, \si{MeV}$  & $-0.028$ &  $0.004$ & $1.028$ & $0.004$ & $43190$ & $2678$ &$2.2$ \\\cmidrule{1-8}
\end{tabular}
\caption{Parameters obtained from fitting the rate-pileup curves of \autoref{fig:Pile_up_vs_rate} with \autoref{eq:pile_up_fit} for both proton energies. Uncertainties on the parameters, denoted by $\delta$, are derived from the errors associated with the fit. $\chi^2_r$ denotes the reduced chi-squared value calculated using the corresponding fit parameters.}
\label{tab:fit_exp_pileup}
\end{table*}
If the particle flux at the TEPC is assumed to be relatively low, $x / \mu \ll 1  $, then the exponential part of \autoref{eq:pile_up_fit} can be approximated to its linear term $\frac{x}{\mu}$.
By imposing a maximum 1\% \replyTo{16}deviation from linearity, the linear approximation is valid for uncorrected particle rates below $1.38 \pm 0.12 \, \unit{kpps}$ at $11 \, \unit{MeV}$ and $0.429 \pm 0.021 \, \unit{kpps}$ for $70\, \unit{MeV}$. 
As the beam intensity increases, both datasets deviate from the linearity, exhibiting a knee and eventually reach a saturation plateau. \\
Both fit parameters of \autoref{tab:fit_exp_pileup} at 11 MeV and 70 MeV are not compatible with the $(0,\,  0)$ point, featuring an offset of $-0.076$ and $-0.024$ for the 11 MeV and 70 MeV energy setup respectively.
The $70\, \si{MeV}$ curve consistently remains above the $11\, \si{MeV}$ curve, indicating a higher pileup for any given rate. This observation is further confirmed by the larger $\mu$ coefficient associated with the $11\, \si{MeV}$ energy, implying a smoother exponential curve, and thus less pileup susceptibility (\autoref{tab:fit_exp_pileup}).

As the particle rate increases, the probability that two distinct electronic signals overlap also increases.
This effect is expected to occur at the stage of the acquisition chain where the signals are slowest, i.e., where their duration is longest, which in the acquisition chain of this study corresponds to the shaping amplifiers.
Signal duration depends on the amplifier settings, more specifically, on the shaping time $t_{\text{shaping}} = 2 \, \si{\micro \s}$.
According to the manufacturer manual, the signal duration is approximately $2.9 \cdot 2.4 \cdot t_{\text{shaping}} = 13.92 \, \si{\micro s} = (71.839 \, \unit{\kilo \Hz})^{-1}$.
Thus, if two or more signals are separated by less than this amount, they will overlap.
This reference value, which is marked in \autoref{fig:Pile_up_vs_rate} in terms of corresponding particle rate and applicable to both energies, serves as a guideline for particle rate matching the inverse of the shaping amplifiers signal duration.
The asymptotic behavior predicted by \autoref{eq:pile_up_fit} in the infinite particle rate limit is consistent in both experimental setups, resulting in an interval compatible with a maximum pileup probability $p = 1$.
According to the proposed model, increasing the pileup probability leads to microdosimetric spectra in which the energy deposition events result from the summation of previously independent events.
Finally, the reduced $\chi^2$ in \autoref{tab:fit_exp_pileup} for both energies suggest that the experimental errors are well estimated, and the overall trend is compatible with \autoref{eq:pile_up_fit} as $\chi^2_r \simeq 1$.

\begin{figure}[htb]
    \centering
    \includegraphics[width=0.7\textwidth]{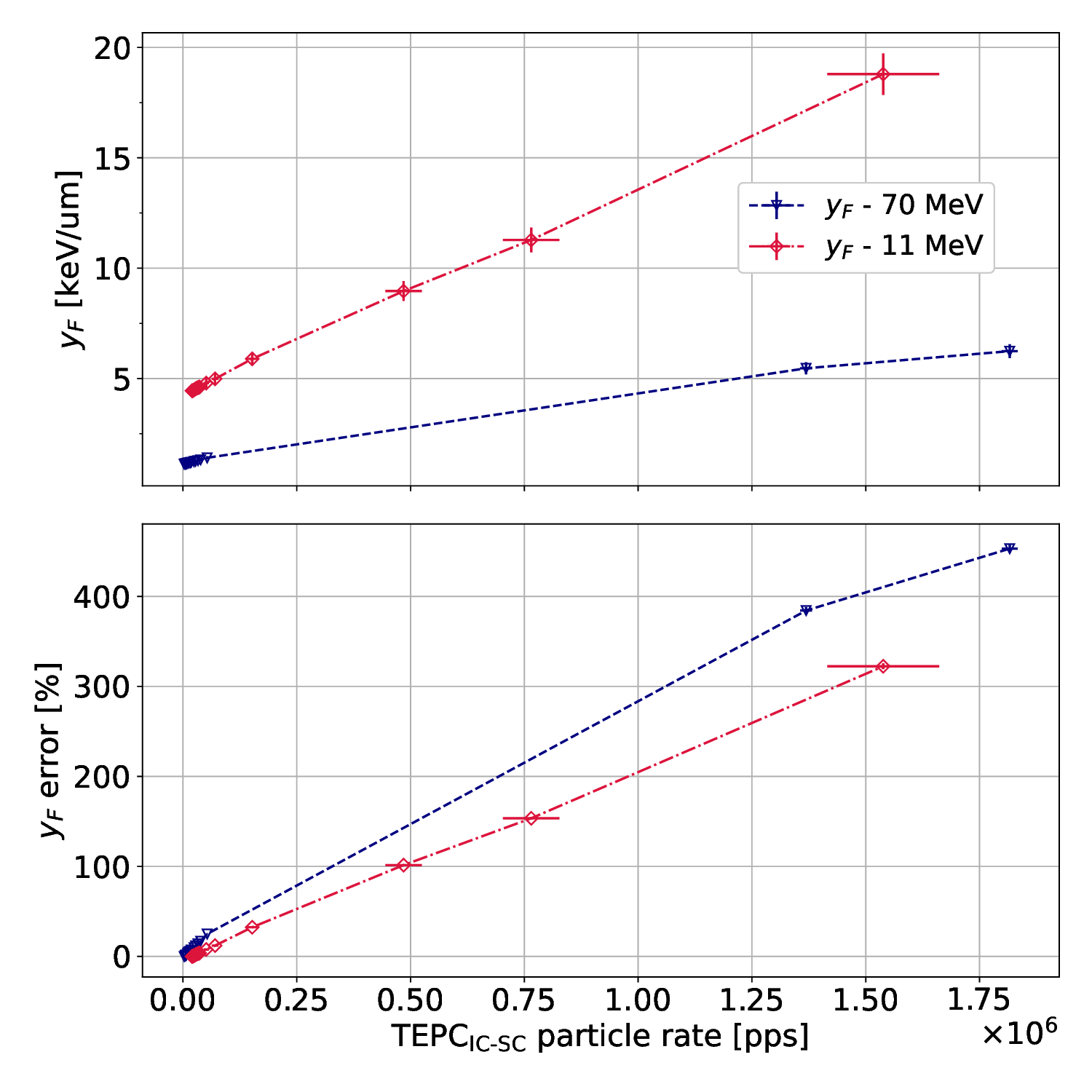}
    \caption{Upper plot:  $y_F$ values measured with the $70 \, \si{MeV}$ (blue triangles) and $11 \, \si{MeV}$ (red diamonds) proton beams.
    Lower plot: percentage errors on the $y_F$ values calculated using the zero pileup $y_F$ as a reference value.
    The dashed lines connecting the points are drawn to guide the reader's eye. 
    }
    \label{fig:y_F_vs_rate}
\end{figure}

To further quantify the impact of pileup on microdosimetric quantities, we evaluated the experimental $y_F$ values at different particle rates \replyTo{1}(TEPC$_{\text{IC} - \text{SC}}$). The $y_F$ least affected by pileup was taken as the reference true value, and the pileup-induced percentage error was then investigated as a function of the TEPC$_{\text{IC} - \text{SC}}$. 
The results are shown in \autoref{fig:y_F_vs_rate} for both energy setups. 
It is possible to observe how the $y_F$ values monotonically increase with the particle rate, with the worst-case scenario showing an increase in $y_F$ by a factor of $\simeq 5$. 
The observed trend in the $y_F$ value can be interpreted as a direct consequence of the pileup effect, as previously described in this section, and as illustrated in \autoref{fig:Pile_up_sample_70MeV}.
\replyTo{2}Additionally, the error associated to the $y_F$ for 70 MeV energy is systematically higher than the same quantity estimated using the 11 MeV energy dataset. 
For this reason, the 70 MeV setup can be considered the worst-case scenario, providing a conservative estimate that can be extended to the 11 MeV energy dataset. Therefore the conclusions from the 70 MeV energy dataset can be extended to the 11 MeV energy dataset.
\replyTo{1}As \autoref{eq:pile_up_fit} in combination with \autoref{tab:fit_exp_pileup} enable to estimate pileup probabilities as a function of (TEPC$_{\text{IC} - \text{SC}}$), it is in turn possible to evaluate the $y_F$ and relative errors as a function of the pileup probability.

\subsection{Pileup estimation based on system dead time}
\label{sec:res:pileup_estimation_DT}
\begin{figure}[htb]
    \centering
    \includegraphics[width=1\textwidth]{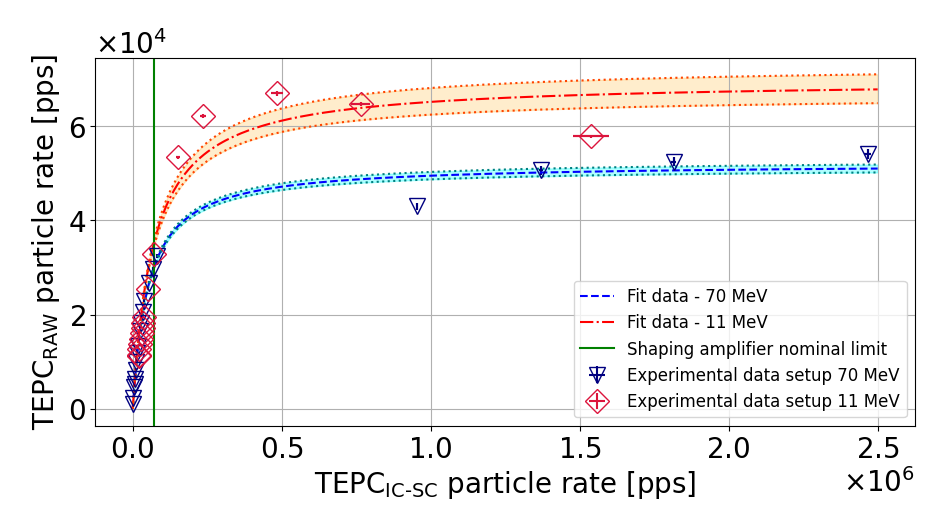}
    \caption{TEPC non-corrected particle rate (TEPC$_{\text{RAW}}$), as a function of the TEPC corrected particle rate (TEPC$_{\text{IC}-\text{SC}}$). \autoref{eq:puleup_counts} is used to fit the data for both explored energies: 11 MeV in red diamonds and 70 MeV in blue triangles. Fit parameters are reported in \autoref{tab:fit_pileup_counts}. The dashed \replyTo{21}curve represents the best fit while the shaded region around each curve denotes the corresponding confidence interval. 
    The vertical green line represents the \replyTo{22}inverse of the typical signal duration according to the shaping amplifier manual of  $\sim 72 \, \unit{\kilo \Hz}$.
    }
    \label{fig:system_response_fit}
\end{figure}

Pileup contribution and associated pileup probability can be estimated using the information from the particle rate measured by the detector without correction (TEPC$_{\text{RAW}}$), and the previously estimated particle rate at the TEPC by means of the IC-SC calibration (TEPC$_{\text{IC}-\text{SC}}$). 
By assuming a non-paralyzable system, it is possible to describe the acquisition system based on the dead time $\tau$ as follows:
\begin{equation}
    m = \frac{n}{n \tau + 1} \iff n = \frac{m}{1 - n\tau}
    \label{eq:puleup_counts}
\end{equation}
where $m$ is the recorded count rate subject to pileup effects (TEPC$_{\text{RAW}}$), and $n$ is the expected count rate without pileup contributions (TEPC$_{\text{IC}-\text{SC}}$).
\autoref{fig:system_response_fit} presents this comparison between the count rate affected by pileup and the expected count rate without pileup contributions. \autoref{eq:puleup_counts} is used to perform the fit \replyTo{19}on the dataset.
 \autoref{fig:system_response_fit} shows how the system can be described as non-paralyzable. 
Similarly to the results from \autoref{fig:Pile_up_vs_rate}, the pileup effect depends on the beam energy, with the \replyTo{20}curve corresponding to the 11 MeV energy setup being consistently above the 70 MeV energy setup.
This behavior can be \replyTo{23}explained by a less severe pileup effect observed in the 11 MeV energy setup.
Additionally, it is possible to observe, in the case of the 11 MeV energy setup, a slowly decreasing trend for the highest particle rates ($>0.5 \, \si{\mega pps}$). This is a typical behavior of a non-ideal system exposed to pileup which is not purely described as a non-paralyzable system, rather a mixture of non-paralyzable and paralyzable.

\autoref{tab:fit_pileup_counts} reports the dead time $\tau$ obtained from the fit of the data in  \autoref{fig:system_response_fit}. 
\begin{table}[htb]
\centering
\scriptsize
\begin{tabular}{lrr}
Energy & $\tau\, [s]$               & $\delta \tau \, [s]$        \\ \hline
11 MeV & $14.3 \times 10^{-6}$ & $0.7  \times 10^{-6}$      \\ 
70 MeV & $19.2 \times 10^{-6}$ & $0.3 \times 10^{-6}$ \\ \hline
\end{tabular}
\caption{Parameters obtained from best fit of \autoref{eq:puleup_counts} on the dataset of \autoref{fig:system_response_fit}. Uncertainties on the $\tau$ parameter are denoted by $\delta \tau$. }
\label{tab:fit_pileup_counts}
\end{table}
Notably, the two dead time estimations are not compatible suggesting an energy dependence. 
Moreover, according to what was observed in the previous section, the 70 MeV energy setup is featuring a higher dead time and therefore, a higher pileup susceptibility if compared to the 11 MeV energy setup. 
Additionally, the system dead time is systematically higher than the nominal signal duration of the shaping amplifier of $2.4 \cdot 2.9 \cdot 2 \, \si{\micro s} = 13.92 \,  \si{\micro s} $ in both cases. 
This result is consistent with the definition of the nominal signal duration, which can represent a theoretical lower limit for the system dead time. In practice, the measured dead time is expected to include contributions from effects that are not solely related to the signal duration, but also arise from other components of the detector and the acquisition chain. These additional contributions degrade the system response at increasing particle rates, effectively increasing the measured dead time.

\begin{figure}[htb]
    \centering
    \includegraphics[width=1\textwidth]{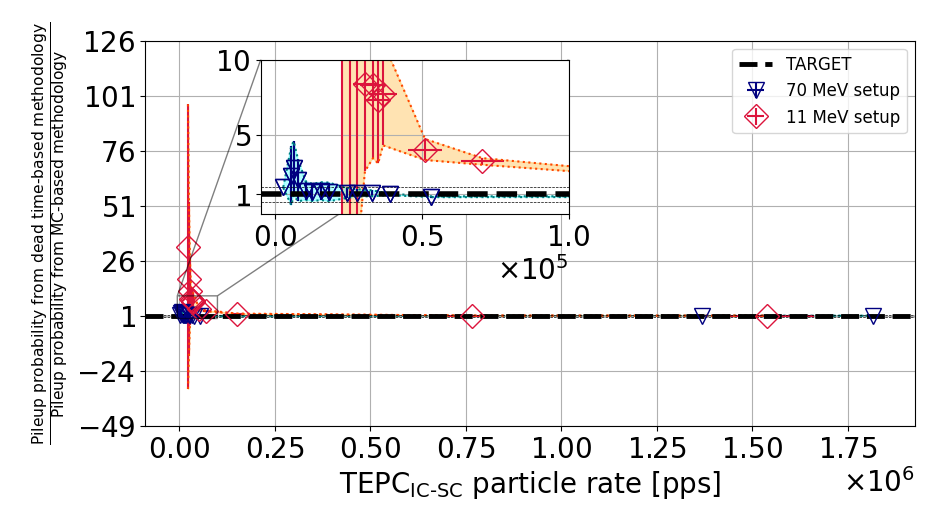}
    \caption{Ratio between the estimated pileup probabilities \replyTo{24}from: $\frac{\text{dead time-based methodology}}{\text{MC-based methodology}} $. Blue triangles represent the 70 MeV energy setup, while red diamonds are used for the 11 MeV energy setup. Lines connect the points to aid readability. The shaded area around the lines represent the confidence interval calculated via error propagation. The horizontal line at $y = 1$ is highlighted.}
    \label{fig:pileup_comparison}
\end{figure}

Combining  \autoref{eq:puleup_counts} with data from \autoref{tab:fit_pileup_counts}, we can calculate a pileup probability as follows:\replyTo{25}
\begin{equation}
 P_{\rm pileup}=1-\exp(-n\tau) \, , 
\end{equation}
with the goal of validating the pileup probabilities derived in \autoref{sec:res:pileup_estimation_MC} using the presented novel methodology.
\autoref{fig:pileup_comparison} presents the ratio between the pileup probabilities: pileup probabilities estimated based on system dead time over the pileup probabilities estimated based on the proposed MC algorithm (\autoref{sec:res:pileup_estimation_MC}).
Ideally, all the points should align on the horizontal line with equation $y = 1$, meaning that the ratio between the two pileup probabilities is unitary. 
This holds true for the points calculated from the 70 MeV energy setup while, for the 11 MeV setup, the ratio shows compatible values within the estimated errors at low ($<30.5 \pm 4.2 \, \si{\kilo pps}$) and high ($>765 \pm 62 \, \si{\kilo pps}$) particle rates. 
This, in turn, corresponds to the low and high limits of the pileup estimation suggesting that the asymptotic behavior is well modeled within the errors.

\section{Discussion and conclusions}
Acquiring microdosimetric data at a clinical rate without pileup effect represents a major experimental challenge. Pileup distorts microdosimetric spectra and their derived quantities, compromising the accuracy of the radiation field characterization. 
To address this problem, we developed an approach that integrates Monte Carlo simulations based on Geant4 with experimental TEPC measurements, enabling us to establish a well-defined relationship between the particle rate at the detector and the corresponding pileup probability. 
We also compared the proposed novel methodology with the theoretical pileup estimation based on counting statistics.
A defining feature of the proposed MC-based pileup estimation method is the use of information embedded in the microdosimetric spectra, while the dead time-based approach relies purely on counting statistics.

According to \autoref{fig:Pile_up_vs_rate} and data from \autoref{tab:fit_exp_pileup}, the pileup probability increases linearly with rate up to approximately $429 \, \unit{pps}$ uncorrected counts, after which it begins to saturate, with the higher beam energy causing a larger pileup effect at any given rate. 
Notably, the pileup probability varies with beam energy. Particularly, \autoref{tab:fit_pileup_counts} and \autoref{tab:fit_exp_pileup} show the same trends, with the 11 MeV energy being less affected by pileup if compared to the 70 MeV energy setup.

Although the fit parameters depend on the beam energy, the two datasets share the same model, which is well described by \autoref{eq:pile_up_fit}. 
Consequently, the values reported in \autoref{tab:fit_exp_pileup} can be used to evaluate a first-order estimate of the pileup probability for experiments performed under similar conditions.\\
Ideally, a comprehensive characterization of the detector response, such as the detailed investigation presented in this work, should be performed to accurately quantify pileup effects. However, practical constraints often make such in-depth analyses unfeasible. In this study, we present a simplified yet validated Monte Carlo–based approach that successfully reproduces the impact of pileup on microdosimetric spectra. Because the method requires only simulated spectra as input, discrepancies observed in experimental spectra acquired under pileup conditions can be associated with specific pileup probabilities, thereby facilitating their interpretation. Although the K-S test was used here to identify the most probable pileup condition for the experimental spectra, alternative metrics could also be employed depending on the application.

\replyTo{1}\replyTo{26}Using the approach outlined in this paper, we can further estimate the pileup-induced error in the microdosimetric $y_F$ values shown in \autoref{fig:y_F_vs_rate}. The results are presented for the 70 MeV energy, which provides a systematically more conservative estimate than the 11 MeV energy due to its greater susceptibility to pileup effects.
An induced error on $y_F$  $<5$ \% can be obtained for both proton energies considering the worst-case scenario of a particle rate of non-corrected TEPC count rate (TEPC$_\text{raw}$) of $10.12 \pm 0.05\,$ kpps, corresponding to a pileup probability of $18 \pm 2 \, \%$. 
\replyTo{27}Similarly, the curve of \autoref{fig:Pile_up_vs_rate} \replyTo{1}can be linearly approximated up to a pileup probability of $0.01 \pm 0.01$. This, in turn, corresponds to a maximum $y_F$ error of $1 \, \%$. 
\replyTo{2}At this particle rate of $429 \pm 21$ pps, the effects of pileup are negligible, ensuring a sufficiently low rate for data acquisition with practically no pileup.

The presented algorithm is based on Poissonian statistics and is designed to account for events with higher multiplicity. Furthermore, the intent is to simulate the different interactions of pileup, such as peak and tail pileup, by performing a weighted sum of the energy deposition.
Although the type of pileup contribution is correlated with the pileup probability, the effect can be modeled with the simplistic assumption of uniform probability.
Using the K–S statistical test as the basis for this approach automatically excludes microdosimetric spectra acquired under extreme conditions, where particle rates greatly exceed the detector capabilities and the physical meaning of the spectra becomes questionable. 

The pileup prediction for the 70 MeV energy setup is in reasonable agreement with the provided error when compared to approaches based on counting statistics. On the other hand, the 11 MeV energy setup features compatible values for a limited range of particle intensities.
Furthermore, both proposed estimates are consistent with an energy-dependent pileup, with the 11 MeV energy being less susceptible to pileup.
Ideally, low-energy protons are supposed to release more energy to the detector, resulting in higher and longer signals in the preamplifier and making them more susceptible to pileup. However, both datasets suggest the opposite with higher pileup effects in configurations with less energy released to the detector.
This effect can be attributed to assumptions shared by both approaches, MC-based and dead time-based, that are not satisfied in practice.
Both approaches are based on Poissonian statistics, and while this assumption holds for radioactive decay sources, it is not representative of particle accelerators such as cyclotrons and synchrotrons \citep{DATA2024169690, knopf2025exploring}. In cyclotrons, particles can be extracted at each radiofrequency (RF) pulse of the machine, resulting in a comb-like temporal structure. 
Typical RF values are around 100 MHz, depending on the machine vendor, implying that the beam is delivered with a comb-like microstructure when the required particle rate approaches the RF frequency, i.e., $\sim 100 \, \si{\MHz}$, corresponding to a period of $10 \, \si{\ns}$. 
Closely time-correlated events are therefore extracted from the machine, violating the Poisson assumption of temporally independent events.
This is likely the case for the 11 MeV beam energy as the particle rate at the cyclotron exit is orders of magnitude higher than the delivered particle rate at the target due to the beam transport and energy selection. Furthermore, the 11 MeV energy setup requires intrinsically higher intensities than the 70 MeV energy setup to compensate for the particle loss due to the RW3 material. 
This, in turn makes the 11 MeV energy setup more susceptible to the beam bunching than the 70 MeV setup, affecting the extracted particle temporal distribution. 
Ultimately, the 11 MeV beam configuration is reproduced less accurately than the 70 MeV configuration by a purely non-paralyzable dead-time model. This observation may suggest that the non-paralyzable dead-time hypothesis, based on Poisson statistics, does not fully describe the system response under the experimental conditions. \\
Several techniques can be used to adjust the particle rate to meet experimental requirements. Their implementation and performance depend on the available resources, operational procedures, and technical capabilities of the specific facility. As a consequence, the fine temporal structure of the extracted beam is inevitably altered, exhibiting characteristic features determined by the adopted beam-control strategy.

In conclusion, the ideal condition for acquiring microdosimetric measurements is, trivially, to operate at particle rates where pileup is negligible. However, this condition is often unattainable and does not reflect realistic scenarios, particularly in clinical environments, where the particle rate cannot be freely adjusted. The methodology presented in this work enables the use of measurements affected by pileup, provided the associated uncertainty remains acceptable, and introduces an approach that allows the microdosimetry community to simulate pileup effects directly on Monte Carlo–generated spectra. Although the method was developed using experimental proton data acquired with a TEPC, the algorithm is readily applicable to other microdosimeters, as well as to different ion species and energies.

\subsection*{Data availability}
All datasets generated and/or analyzed during this study are available from the corresponding authors on reasonable request.

\bibliographystyle{elsarticle-num} 
\bibliography{sample}

\end{document}